\documentclass[aps,prl,twocolumn,preprintnumbers,amssymb,nobibnotes,nofootinbib,longbibliography,superscriptaddress]{revtex4-2}
\usepackage[utf8]{inputenc}
\usepackage[T1]{fontenc}

\usepackage{amsmath, amsfonts, amssymb}
\usepackage{mathtools}
\usepackage{physics}
\usepackage{graphicx}
\usepackage{svg}
\usepackage{booktabs}
\usepackage{array}
\usepackage{enumitem}
\usepackage{hyperref}

\hypersetup{%
    colorlinks=true,%
    allcolors={blue!60!black}%
}

\newcommand{\CS}{\textcolor{blue!60!black}}
\newcommand{\nf}{n_f}

\allowdisplaybreaks
\usepackage{cleveref}

\graphicspath{ {figures/} }
\newcommand{\FDiag}[2]{
\begin{minipage}{0.155\textwidth}
\begin{center}
\CS{#1}\\[-2ex]
\includegraphics[angle=-90,width=\textwidth]{#2}
\end{center}
\end{minipage}
\hspace*{-2ex}
}

\begin{document}

\title{The four-loop non-singlet splitting functions in QCD}

\author{Thomas~Gehrmann}
\email{thomas.gehrmann@uzh.ch}
\affiliation{Physik-Institut, Universit\"at Z\"urich, Winterthurerstrasse 190, 8057 Z\"urich, Switzerland}

\author{Andreas von Manteuffel}
\email{manteuffel@ur.de}
\affiliation{Institut f\"ur Theoretische Physik, Universit\"at Regensburg, 93040 Regensburg, Germany}

\author{Vasily Sotnikov}
\email{vasily.sotnikov@physik.uzh.ch}
\affiliation{Physik-Institut, Universit\"at Z\"urich, Winterthurerstrasse 190, 8057 Z\"urich, Switzerland}

\author{Tong-Zhi Yang}
\email{tongzhi.yang@m.scnu.edu.cn}
\affiliation{State Key Laboratory of Nuclear Physics and Technology, Institute of Quantum Matter,
South China Normal University, Guangzhou 510006, China}
\affiliation{%
Guangdong Basic Research Center of Excellence for Structure and Fundamental Interactions of Matter, Guangdong Provincial Key Laboratory of Nuclear Science, Guangzhou 510006, China
}

\date{\today}

\begin{abstract}
The scale evolution of parton distributions is governed by splitting functions.  
We compute the four-loop splitting functions in perturbative QCD that control the evolution of quark non-singlet distributions.
We confirm previous partial results and obtain, for the first time, fully analytic expressions for all non-singlet contributions at this order. 
These allow us to extract the analytic form of the four-loop virtual 
and rapidity anomalous dimensions entering logarithmic resummation.
We provide precise numerical representations of the splitting functions suitable for parton evolution.
\end{abstract}

\maketitle

\preprint{ZU-TH 15/26}

\section{Introduction}
The proton is a complex bound state of quarks and gluons. Its 
inner structure is encoded in parton distributions functions (PDFs), 
which describe~\cite{Bjorken:1968dy,Bjorken:1969ja} the probability of finding a quark, anti-quark or gluon 
carrying a fraction $x$ of the proton's momentum when probing it at 
a resolution scale $Q^2$. The PDFs encode the non-perturbative bound state 
dynamics of the proton,  
they are universal quantities appearing in the calculation of 
any collider process involving hadrons in the initial state. 
Their $Q^2$-evolution is predicted in perturbative 
QCD by the DGLAP equations~\cite{Altarelli:1977zs,Gribov:1972ri,Dokshitzer:1977sg}, 
which form of a coupled set 
of integro-differential equations. The kernels of 
the DGLAP equations are the splitting functions $P(x)$, which determine 
the probability of partons transitioning into each other while transferring 
a given fraction $x$ of their longitudinal momentum. 

The splitting functions can be computed in a loop expansion in QCD
perturbation theory, and their knowledge to a given order enables 
the extraction of PDFs accurate to this order from a global fit 
to experimental data from a wealth of processes in lepton-hadron and 
hadron-hadron collisions. Analytical expressions for the 
splitting functions were derived at one loop~\cite{Gross:1974cs,Politzer:1974fr,Altarelli:1977zs},
two loops~\cite{Floratos:1977au,Floratos:1978ny,Gonzalez-Arroyo:1979guc,Curci:1980uw,Furmanski:1980cm,Hamberg:1991qt}
and three loops~\cite{Moch:2004pa,Vogt:2004mw,Blumlein:2021enk,Blumlein:2021ryt,Gehrmann:2023ksf},
enabling PDF fits at LO, NLO and NNLO, respectively. 
At four loops, approximate 
expressions for the splitting functions~\cite{Moch:2017uml,Moch:2021qrk,Falcioni:2023luc,Falcioni:2023vqq,Falcioni:2024xyt,Falcioni:2024qpd,Falcioni:2025hfz}
as well as partial analytical results~\cite{Gracey:1994nn,Davies:2016jie,Gehrmann:2023cqm,Gehrmann:2023iah,Falcioni:2023tzp,Kniehl:2025ttz,Kniehl:2025jfs} 
are available, which 
are already being used in PDF determinations at N\textsuperscript{3}LO \cite{McGowan:2022nag,Cooper-Sarkar:2024crx,NNPDF:2024nan,Cridge:2024icl,Hampson:2025pvi,Karlberg:2025hxk}. 
The 
approximate four-loop splitting functions may be sufficient for many 
phenomenological applications. Knowledge of 
their complete analytical expressions is nevertheless highly desirable, since 
it will enable to eliminate the approximation uncertainty and 
allows for in-depth analytical 
studies of their asymptotic behaviour at large and 
small $x$. 

For $\nf$ quark flavours, the $(2\nf+1)$ DGLAP evolution equations can be 
decoupled by introducing $(2\nf-2)$ flavour-asymmetric non-singlet 
combinations of the form
\begin{equation}
    q_{\mathrm{ns},ij}^\pm = q_i\pm \bar{q}_i - (q_j\pm \bar{q}_j)
\end{equation}
and one non-singlet valence sum distribution
\begin{equation}
    q_{\mathrm{ns}}^V = \sum_i (q_i- \bar{q}_i)\;,
\end{equation}
thus leaving a system of 
two coupled equations for the quark singlet and 
the gluon. In this letter, we compute analytical expressions for 
the three different 
four-loop splitting functions $P_{\mathrm{ns}}^{\pm,V}$,
with $P_{\mathrm{ns}}^{V} = P_{\mathrm{ns}}^{-}+
P_\mathrm{ns}^{s}$,
that govern the evolution of the 
different non-singlet quark distributions. 
These space-like splitting functions directly determine~\cite{Dokshitzer:2005bf,Mitov:2006ic,Basso:2006nk,Moch:2017uml} their time-like counterparts, which describe the evolution of non-singlet hadron fragmentation functions.

\section{Methodology}

\begin{figure*}[]
\begin{center}
 \FDiag{$C_F C_A^3$}{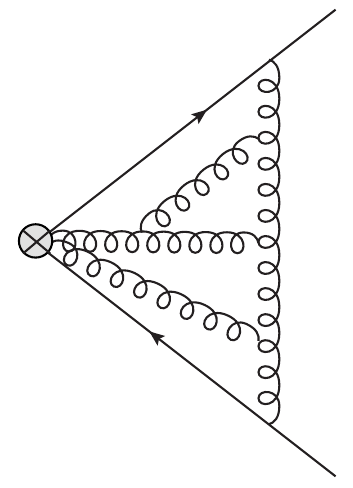}
 \FDiag{$C_F^2 C_A^2$}{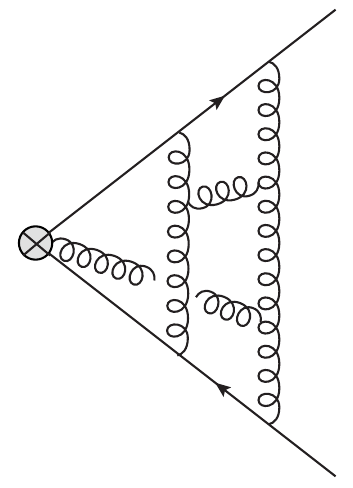}
 \FDiag{$C_F^3 C_A$}{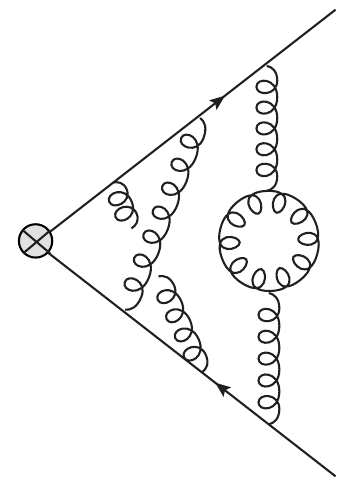}
 \FDiag{$C_F^4$}{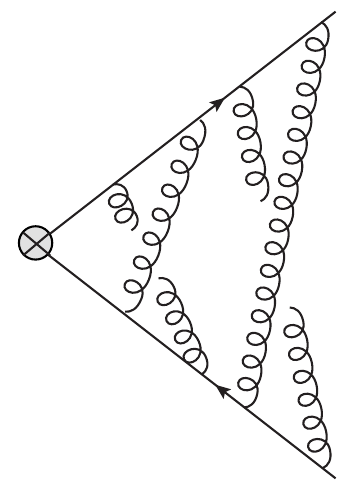}
 \FDiag{$d_F^{abcd}d_A^{abcd}/N_c$}{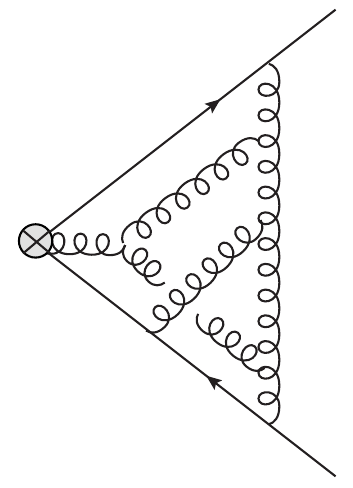}
 \FDiag{$\nf (d_F^{abcd})^2/N_c$}{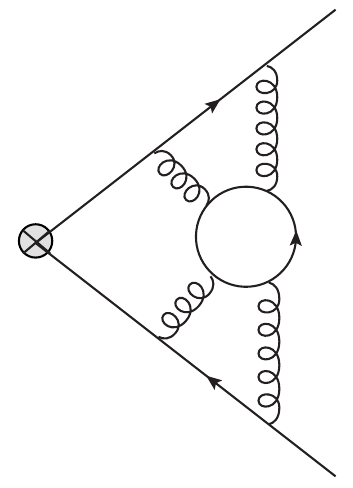}
\\[1ex]
 \FDiag{$\nf C_F C_A^2$}{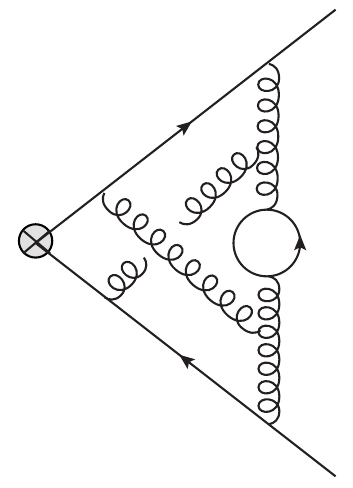}
 \FDiag{$\nf C_F^2 C_A$}{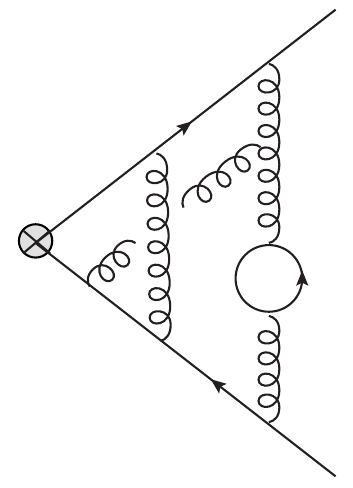}
 \FDiag{$\nf C_F^3$}{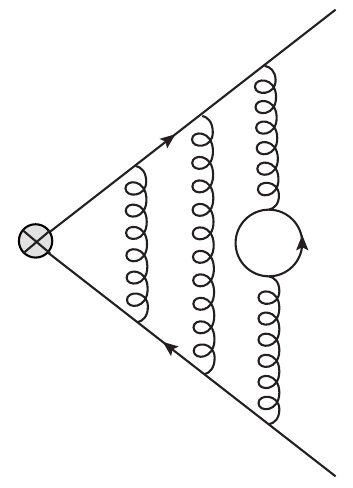}
 \FDiag{$\nf^2 C_F C_A$}{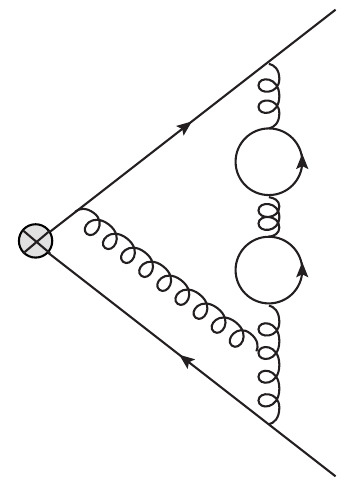}
 \FDiag{$\nf^2 C_F^2$}{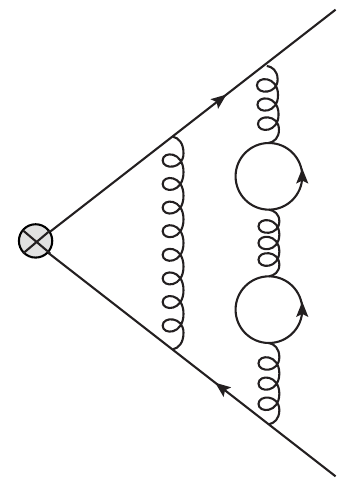}
 \FDiag{$\nf^3 C_F$}{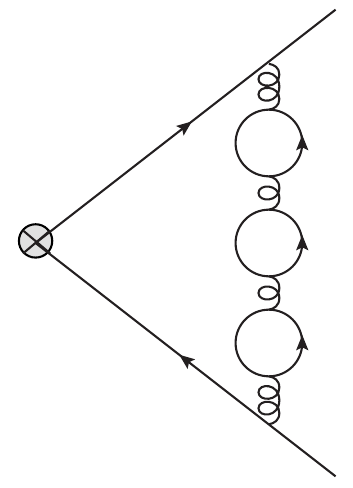}
\\[1ex]
 \FDiag{$\nf (d_F^{abc})^2 C_A / N_c$}{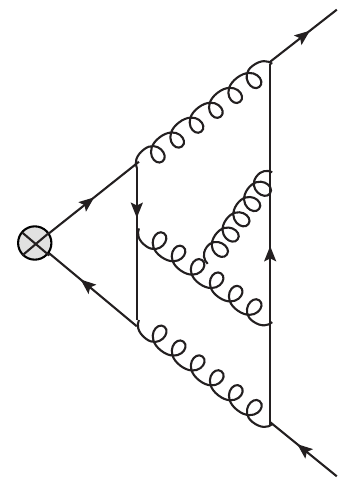}
 \FDiag{$\nf (d_F^{abc})^2 C_F / N_c$}{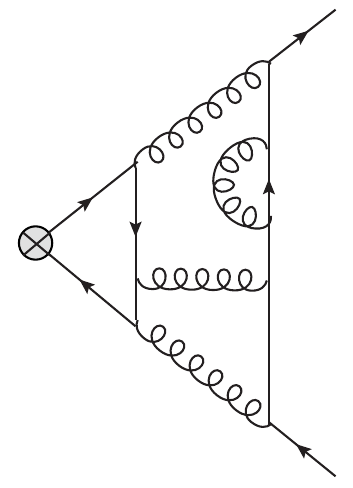}
 \FDiag{$\nf^2 (d_F^{abc})^2 / N_c$}{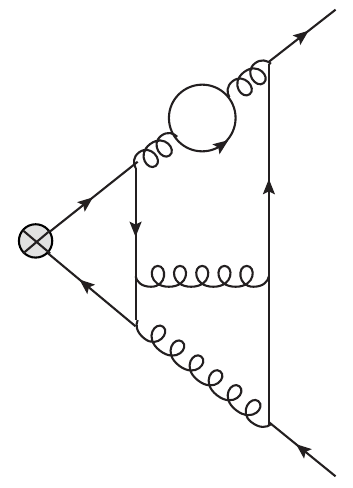}
\end{center}
\caption{\label{fig:diags}Sample Feynman diagrams for off-shell quark self energies with an operator insertion (crossed vertex) contributing to different color coefficients of the splitting functions $P_\mathrm{ns}^{(3) \pm}$ (upper two rows) and $P_\mathrm{ns}^{(3) s}$ (lower row) at four loops.
Both, planar and non-planar diagrams contribute.
The operator insertion may involve up to four gluons (curly lines) in addition to the two quarks (straight lines).
}
\end{figure*}

The QCD splitting functions $P(x)$ relate to the anomalous dimensions $\gamma(n)$ of quark and gluon operators through the Mellin transformation:
\begin{equation}
\label{eq:MellinT}
\gamma(n) =  - \int_0^1 d x\, x^{n-1} P(x)\,.
\end{equation} 
The splitting functions can therefore be extracted from 
the anomalous dimensions $\gamma(n)$ for symbolic Mellin moment $n$.
These are determined in the operator product expansion from the 
the respective two-parton operator matrix elements (OME) in off-shell kinematics~\cite{Gross:1974cs,Politzer:1974fr}. 
In the non-singlet case we compute
\begin{align}
\label{eq:OMEsDe}
A^{}_{\mathrm{ns}} = \mel{q(p)}{~O_{\mathrm{ns}}~}{q(p)} \text{ with } p^2<0\,.
\end{align}
The twist-two operator of spin $n$ 
\begin{equation}
\label{eq:nonsingletOP}
O^{}_{\mathrm{ns}}(n) = \frac{i^{n-1}}{2}  \bar{\psi} \,\Delta\cdot \gamma \,(\Delta \cdot D)^{n-1}\,\frac{\lambda}{2} \psi  
\end{equation}
contains the flavor space matrices $\lambda$ which project onto the different non-singlet quark combinations.
Here, covariant derivatives $D_\mu$ and Dirac matrices $\gamma_\mu$
are contracted with a light-like auxiliary vector $\Delta_\mu$ to
create a scalar operator.
 
The renormalization of the non-singlet operators takes a 
multiplicative form
\begin{equation}
\label{eq:nsRenormalization}
O^{\text{R}}_{\text{ns}}(\mu, n)  = Z_{\text{ns}}(\mu, n) O^{\text{B}}_{\text{ns}}(n) \,,
\end{equation}
with $O^{\text{B}}$ and $O^{\text{R}}$ 
denoting the bare and renormalized operators, respectively. 
This simple multiplicative renormalization is in stark contrast to 
the quark singlet sector, which mixes with the gluon operator and 
with further gauge-variant operators, leading to a conceptually involved 
renormalization pattern~\cite{Gross:1974cs,Dixon:1974ss,Hamberg:1991qt} that 
could be fully understood only quite recently~\cite{Falcioni:2022fdm,Gehrmann:2023ksf,Falcioni:2024xav,Gehrmann:2024ggw}.

The anomalous dimension $\gamma_\mathrm{ns}$ is defined
through
\begin{equation}
\label{eq:anomdimdef}
 \frac{d Z_{\text{ns}}(\mu,n) }{d \ln \mu} = -2  \gamma_{ \text{ns}}(\mu,n) \, Z_{\text{ns}}(\mu,n) \,. 
\end{equation}
Its perturbative expansion in the renormalized coupling $a_s=\alpha_s(\mu)/(4\pi)$ reads
\begin{equation}
\gamma_{\text{ns}}  =  \sum_{l=0}^\infty a_s^{l+1} \gamma_{\text{ns}}^{(l)}\,,
\end{equation}
where $\gamma_{\text{ns}}^{(l)}$ is the $(l+1)$ loop 
contribution.
The dependence on the Mellin moment index $n$ is omitted for brevity. 
Employing dimensional regularization in $4-2\epsilon$ dimensions, one can extract the anomalous dimensions
from the single 
$\epsilon$-pole terms of renormalization constants $Z_\mathrm{ns}(\mu,n)$.

At one-loop order, all three non-singlet anomalous dimensions agree  
with each other. $\gamma_{\text{ns}}^-$ and $\gamma_{\text{ns}}^+$ 
differ from two loops onwards
and $\gamma_{\text{ns}}^V$ deviates from $\gamma_{\text{ns}}^-$ starting at 
three loops. 

To compute the four-loop anomalous dimensions, we calculate Feynman
diagrams for matrix elements \eqref{eq:OMEsDe} with two external quarks
and an operator insertion on a quark line.
We generate $\mathcal{O}(16\mathrm{k})$
diagrams with \texttt{Qgraf} \cite{Nogueira:1991ex}, from which
we select the ones contributing to the non-singlet anomalous dimensions.
Example diagrams are shown in \cref{fig:diags}.

We perform the Lorentz, Dirac and color algebra with
\texttt{Form} \cite{Vermaseren:2000nd,Davies:2026cci} and \texttt{Color.h} \cite{vanRitbergen:1998pn}.
For an operator of spin $n$ the corresponding Feynman rule gives rise to terms like
$(\Delta \cdot k)^{n-1}$, where $k$ is the momentum
of any particle adjacent to the operator. 
These symbolic powers in numerator insertions can be 
transformed into linear propagators by the introduction of an auxiliary parameter 
$t$~\cite{Ablinger:2012qm,Ablinger:2014nga}. 
For example, we replace
\begin{equation}
    (\Delta \cdot k)^{n-1} 
    \rightarrow \sum_{n=1}^\infty t^n (\Delta \cdot k)^{n-1} = \frac{t}{1-t \Delta \cdot k},
\end{equation}
thus allowing to 
perform symbolic manipulations and reexpand the final result to extract
the contribution for a specific value $n$.

The resulting scalar Feynman integrals can be indexed by 52 complete sets of standard and
linear propagators (integral families) and assigned to 549 top-level sectors with 13 different propagators and their subsectors.
Here, a sector refers to a set of different propagator denominators.
We encounter $\mathcal{O}(3\mathrm{M})$ unreduced integrals  with at most 13 different propagators.
Using momentum shift relations, this number reduces to $\mathcal{O}(260\mathrm{k})$
integrals,
with up to 5 irreducible scalar products in the numerator and up to 4 repeated propagators
(``dots'').
We employ integration-by-parts (IBP) reductions
\cite{Chetyrkin:1980pr,Laporta:2000dsw}
with \texttt{Reduze\;2} \cite{vonManteuffel:2012np}
and the private code \texttt{Finred}.
Standard IBP relations reduce the matrix elements in terms
of $\mathcal{O}(120\mathrm{k})$ integrals.
Taking also shift relations between sectors, discrete symmetry relations within sectors,
and certain anomalous IBP identities generated in supersectors into account,
one can express all integrals in terms of $\mathcal{O}(6\mathrm{k})$
master integrals.

For the matrix element reduction, we employ standard momentum-space IBP
relations with finite field sampling
\cite{vonManteuffel:2014ixa,Peraro:2016wsq}
and a custom implementation to optimize the selection of equations,
see also \cite{Driesse:2024xad,Guan:2024byi,Bern:2024adl,vonHippel:2025okr,Song:2025pwy,Lange:2025fba}.
For a given top-level sector, we determine a set of spanning sectors
\cite{Larsen:2015ped,Guan:2024byi}
and sample the reduced matrix element for each of them.
In total, we probe $\mathcal{O}(9\mathrm{k})$
spanning sectors.
We make an effort to avoid the introduction of denominator factors
with mixed $\epsilon$ and $t$ dependence through the choice of
the basis integrals~\cite{Smirnov:2020quc,Usovitsch:2020jrk}.
After guessing denominators and simple numerator factors
\cite{Abreu:2018zmy,Heller:2021qkz}, we reconstruct
the reduced matrix element from finite field
samples.

To find the solutions of the basis integrals we derive
differential equations (DEs)~\cite{Gehrmann:1999as,Henn:2013pwa}
with respect to the tracing parameter $t$ for $\order{300}$ top-level sectors.
For each top-level sector $\tau$ we obtain
\begin{equation} \label{eq:DE}
  \partial_t \, I^{\tau} ~=~ A^{\tau}(\epsilon, t) \, I^{\tau}\,,
\end{equation}
where the vector $I^\tau$ contains the basis integrals
and $A^\tau(\epsilon, t)$ is a matrix.
A fifth of all DEs have dimension larger than 900, with the two largest ones exceeding 2000, and homogeneous coupled blocks of size 12 are commonly encountered.

We derive Laurent series solutions in $\epsilon$ with coefficients that are Taylor series in $t$ as follows.
From the  DEs we derive recursion relations for series coefficients of each integral \cite{Blumlein:2009tj, Ablinger:2015tua, Lee:2017qql}.
The solutions of interest to us are required to be regular in the
limit $t\to 0$.
Moreover, in the limit $t \to 0$ all integrals are reduced to a
basis of 28 standard self-energy integrals
\cite{Baikov:2010hf,Lee:2011jt,vonManteuffel:2019gpr,Lee:2023dtc},
which fix all boundary conditions.
We expand the boundary integrals to sufficiently high orders in $\epsilon$ such that contributions up to weight 7 to the OMEs are fully captured.

We use the recursion relations to obtain a sufficient number of $t$ powers for all integrals that allows us to compute the four-loop bare OMEs up to $\mathcal{O}(4000)$ moments. From these moments, we successfully reconstruct the four-loop anomalous dimensions $\gamma^{(3)}_{\text{ns}}$ in terms of harmonic sums~\cite{Vermaseren:1998uu}, employing the renormalization formula~\eqref{eq:nsRenormalization} and lower-order OME counterterms~\cite{Gehrmann:2023ksf}. Unlike the strong constraints imposed in~\cite{Moch:2017uml,Kniehl:2025ttz}, our $n$-space reconstruction uses a generic ansatz: for $\gamma^{(3)\pm}_{\text{ns}}$, harmonic sums up to weight 7 and two denominator structures $\{n,\,n+1\}$; for $\gamma^{(3)s}_{\text{ns}}$, harmonic sums up to weight 6 and five denominator structures $\{n,\,n+1,\, n-1,\,n+2,\,n-2\}$.

\begin{figure}[t]
  \centering
  \includegraphics[width=0.4\linewidth]{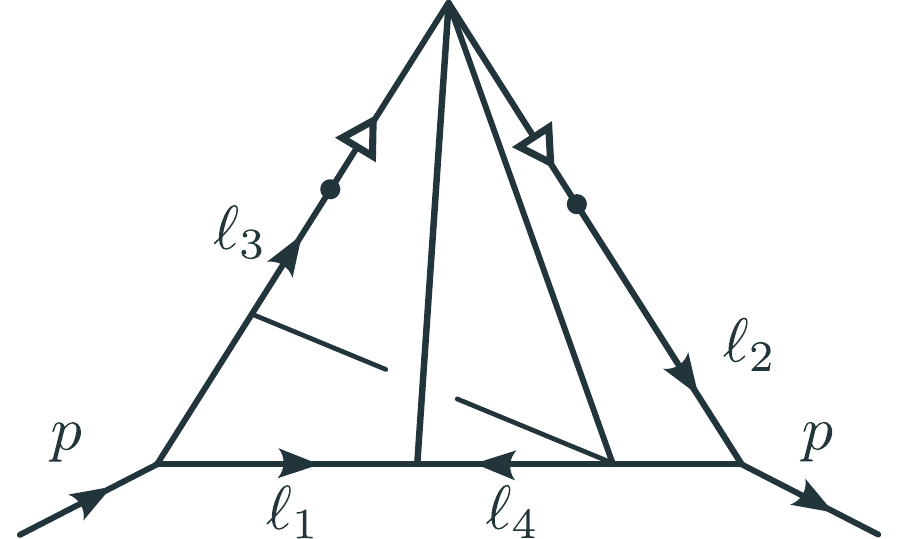}
  \caption{Feynman integral topology with 11 denominators associated to elliptic geometry.
  This topology occurs as a subtopology of
  the fifth diagram in \cref{fig:diags}.
  The lines marked with hollow triangles and dots correspond to insertions of denominators $1- t\,(\Delta \cdot \ell_i)$ with tracing parameter $t$ in addition to standard (quadratic) propagators.}
  \label{fig:elliptic-sector}
\end{figure}

While our method of DE solution is not particularly concerned with the structure of rational matrices $A^{\tau}(\epsilon, t)$, 
we studied the solutions of their homogeneous blocks $\hat{A}^{\tau}_i(\epsilon, t)$ in more detail for some examples.
In addition to square root solutions that do not appear in a three-loop computation, we also surprisingly find that integrals related to elliptic geometry appear at four loops for the first time.
For example, for the integrals with denominators shown in \cref{fig:elliptic-sector} we initially find a $4\times4$ homogeneous coupled DE $B_e = \hat{A}^{\tau}_e(0, t)$.
Using a private extension of the package \texttt{DLogBasis} \cite{Henn:2020lye} we obtain two candidates which have $\dd\log$ integrands in four-dimensional momentum parametrization which decouple from $B_e$.
We convert the remaining $2\times 2$ system into a second-order DE whose solutions are periods of an elliptic curve. We then use the ansatz method of \cite{Bogner:2019lfa} to find an $\epsilon$-factorized form with a transformation 
that contains complete elliptic integrals of the first and second kind. 
Interestingly, if one were to derive exact solutions in $t$, instead of the Taylor series solutions here, 
the iterated integrals with elliptic kernels originating from such Feynman integrals are expected to completely drop out from the poles in $\epsilon$ of the bare OME result,
since we are able to recover the all-$n$ result expressed through harmonic sums.

\section{Results}

Our analytic results for $\gamma^{(3)}_{\mathrm{ns}}$ are expressed solely
in terms of harmonic sums and decomposed into 15 color structures, 
as shown in fig.~\ref{fig:diags}, with
\begin{gather}
\label{eq:casimir}
T=1/2,~
C_A = 3,~
C_F = 4/3,~
(d_F^{abc})^2/N_c = 5/18,\nonumber\\
d_F^{abcd} d_A^{abcd}/N_c = 5/2,~
(d_F^{abcd})^2/N_c = 5/36\,.
\end{gather}
for the $SU(3)$ gauge group of QCD.
The non-fermionic contributions 
to $\gamma^{(3)\pm}_{\mathrm{ns}}$ and the $\nf$ term in 
$\gamma^{(3)s}_{\mathrm{ns}}$, represented by the first five 
structures in the top row and the first two in the bottom row of 
fig.~\ref{fig:diags}, are obtained here for the first time. Analytic results for other color structures (including the leading color) were initially derived 
in~\cite{Gracey:1994nn,Davies:2016jie,Moch:2017uml,Gehrmann:2023iah,Kniehl:2025ttz} and are confirmed 
here. 
 By evaluating our results with fixed $n$, we found full agreement with fixed moment results in~\cite{Moch:2017uml}. We observed that the harmonic sums of weight 6 and denominator structure $\{n-2\}$ in $n$-space ansatz drop out for $\gamma^{(3)s}_{\text{ns}}$, however the result does contain terms like $S_{1,1,-3}(n)/n^2$, which is weight 7 if we count the $1/n^2$ as weight 2.

By performing an inverse Mellin transformation with respect to eq.~\eqref{eq:MellinT}, as implemented in the \texttt{HarmonicSums} package~\cite{Ablinger:2009ovq,Ablinger:2012ufz}, the corresponding analytic four-loop non-singlet splitting functions are obtained. The results contain harmonic polylogarithms~\cite{Remiddi:1999ew,Gehrmann:2001pz} up to weight 6, and transcendental constants up to weight 7, appearing in the $\delta(1-x)$ part of $P^{(3)\pm}_{\text{ns}}$.
For numerical evaluations and expansions of the harmonic
polylogarithms, we employ the package \texttt{HPL} \cite{Maitre:2005uu}
and the implementation \cite{Vollinga:2004sn} of $G$ functions in \texttt{GiNaC}.

Our analytic results allow us to study the asymptotic behaviour of the splitting functions. In the limit of $x \to 1$, $P^{(3)s}_{\text{ns}}$ vanishes at leading and next-to-leading power. The $P^{(3) +}_{\text{ns}}$ and $P^{(3) -}_{\text{ns}}$ are identical to next-to-leading power, and can be written as
\begin{align}
\label{eq:xto1limit}
    P^{(3) \pm}_{\text{ns}}=& A_4 \left[\frac{1}{1-x}\right]_+ +   B_4 \, \delta(1-x)  \nonumber \\
    &+  C_{4} \log (1-x)+  D_{4} - A_4 + \mathcal{O}(1-x)\,,
\end{align}
where $[~]_+$ denotes the plus distribution \cite{Altarelli:1977zs,Floratos:1981hs}. 
The $A_4$ is the four-loop cusp anomalous dimension~\cite{Korchemsky:1987wg}, derived in~\cite{Henn:2019swt,vonManteuffel:2020vjv}, where we find full agreement. The $B_4$ is called four-loop virtual anomalous dimension, 
and we obtain its analytic form here for the first time:
\begin{align}
 &B_4 =  \CS{C_A^3 C_F} \biggl( 
  - \frac{8960 }{3}\zeta_7
  + \frac{1472}{3} \zeta_5 \zeta_2
  + \frac{32}{3} \zeta_4 \zeta_3
  + \frac{73333}{108} \zeta_6
  \nonumber\\ &
  + \frac{1672}{3} \zeta_3^2
  + \frac{11522 }{9}\zeta_5
  + \frac{584}{3} \zeta_3 \zeta_2
  - \frac{11206}{27} \zeta_4
  - \frac{152284 }{81}\zeta_3
  \nonumber\\ &
  + \frac{13864}{9} \zeta_2
  - \frac{373793}{648}
  \biggr)
 + \CS{C_A^2 C_F^2} \biggl( 
   8610 \zeta_7
  - 2104 \zeta_5 \zeta_2
  \nonumber\\ &
  - 32 \zeta_4 \zeta_3
  - \frac{5497}{2}\zeta_6
  - \frac{7102 }{3}\zeta_3^2
  + \frac{5354}{9} \zeta_5
  + \frac{2096 }{9}\zeta_3 \zeta_2
  \nonumber\\ &
  - \frac{60850 }{27}\zeta_4
  + \frac{129662 }{27}\zeta_3
  - \frac{46771 }{27}\zeta_2
  + \frac{29639}{36}
  \biggr) \nonumber \\
&+ \CS{C_A C_F^3} \biggl( 
  - 10920 \zeta_7
  + 2064 \zeta_5 \zeta_2
  + 128 \zeta_4 \zeta_3
  + \frac{79297}{18}\zeta_6
  \nonumber\\ &
  + 3220 \zeta_3^2
  - 976 \zeta_5
  - \frac{1988 }{3}\zeta_3 \zeta_2
  + 2167\zeta_4
  - 3260 \zeta_3 
  \nonumber\\ &
  + 1167 \zeta_2
  - \frac{2085}{4}
  \biggr)
+ \CS{C_F^4} \biggl( 
   5880 \zeta_7
  - 384 \zeta_5 \zeta_2
  \nonumber\\ &
  + 64 \zeta_4 \zeta_3
  -2111 \zeta_6
  - 1152 \zeta_3^2
  - 2520 \zeta_5
  - 120 \zeta_3 \zeta_2
  - 342 \zeta_4
  \nonumber\\ &
  + 2004 \zeta_3
  - 450 \zeta_2
  + \frac{4873}{24}
  \biggr)
  + \CS{\frac{d_F^{abcd} d_A^{abcd} }{N_c} } \biggl( 
   2800 \zeta_7
  \nonumber\\ &
  + 320 \zeta_5 \zeta_2
  - 64 \zeta_3 \zeta_4
  -\frac{1562}{9}\zeta_6
  - 704 \zeta_3^2
  - 400 \zeta_5 
  - 896 \zeta_3 \zeta_2
  \nonumber\\ &
  + \frac{32}{3} \zeta_4
  - \frac{1232 }{3}\zeta_3
  - \frac{944 }{3}\zeta_2
  + 96 
  \biggr)
+ \CS{C_A^2 C_F \nf} \biggl( 
  -\frac{3913}{27} \zeta_6
  \nonumber\\ &
  + \frac{416 }{3}\zeta_3^2
  + \frac{1130 }{9}\zeta_5
  - \frac{580 }{3}\zeta_3 \zeta_2
  + \frac{1234 }{9}\zeta_4
  + \frac{9554 }{27}\zeta_3
  \nonumber\\ &
  - \frac{41092 }{81}\zeta_2
  + \frac{20027}{108}
  \biggr)
+ \CS{C_A C_F^2 \nf} \biggl( 
   351\zeta_6
  - \frac{1232 }{3}\zeta_3^2
  \nonumber\\ &
  - \frac{7432 }{9}\zeta_5
  + \frac{2672 }{9}\zeta_3 \zeta_2
  + \frac{27854}{27}\zeta_4
  - \frac{15400 }{27}\zeta_3
  - \frac{3892 }{27}\zeta_2
  \nonumber\\ &
  - \frac{7751}{54}
  \biggr)
+ \CS{C_F^3 \nf} \biggl( 
  -\frac{6434 }{9}\zeta_6
  + 224 \zeta_3^2
  + 912 \zeta_5
  - \frac{256 }{3}\zeta_3 \zeta_2
  \nonumber\\ &
  - 204 \zeta_4
  - 308 \zeta_3
  + 162 \zeta_2
  + 32
  \biggr)
+ \CS{\frac{(d_F^{abcd})^2 \, \nf}{N_c}} \biggl( 
   \frac{1792}{9}\zeta_6
  \nonumber\\ &
  + 256 \zeta_3^2
  - 1120 \zeta_5
  + 64 \zeta_3 \zeta_2 
  - \frac{352}{3} \zeta_4
  - \frac{992}{3} \zeta_3
  + \frac{1888}{3} \zeta_2
  \nonumber\\ &
  - 192
  \biggr)
+ \CS{C_A C_F \nf^2} \biggl( 
  - \frac{88 }{9}\zeta_5
  + \frac{80 }{3}\zeta_3 \zeta_2
  -\frac{80}{9}\zeta_4
  - \frac{320 }{9}\zeta_3
  \nonumber\\ &
  + \frac{3170 }{81}\zeta_2
  - \frac{193}{54}
  \biggr)
+ \CS{C_F^2 \nf^2} \biggl( 
   \frac{368 }{9}\zeta_5
  - \frac{160 }{9}\zeta_3 \zeta_2
  -\frac{2104 }{27}\zeta_4
  \nonumber\\ &
  + \frac{56 }{27}\zeta_3
  + \frac{1244 }{27}\zeta_2
  - \frac{188}{27}
  \biggr)
+ \CS{C_F \nf^3} \biggl( 
  -\frac{32}{27}\zeta_4
  + \frac{304 }{81}\zeta_3
  \nonumber\\ &
  + \frac{32 }{81}\zeta_2
  - \frac{131}{81}
  \biggr).
\end{align}
Approximate numerical results for $B_4$ have been presented in~\cite{Das:2019btv}, 
they are in good agreement with the analytical result above. 
Together with the four-loop collinear anomalous dimensions in~\cite{Agarwal:2021zft,vonManteuffel:2020vjv}, our result for $B_4$ analytically determines the soft anomalous dimension at four loops.
The soft-rapidity correspondence derived in~\cite{Li:2016ctv,Vladimirov:2016dll,Vladimirov:2017ksc} relates this quantity to the rapidity anomalous dimension.
Our calculation of $B_4$ promotes a previous determination \cite{Moult:2022xzt,Duhr:2022yyp} of the four-loop rapidity anomalous dimension in terms of numerical constants to a fully analytical result, documented in the appendix.

\begin{figure*}[ht]
  \centering
  \includegraphics[width=0.49\linewidth]{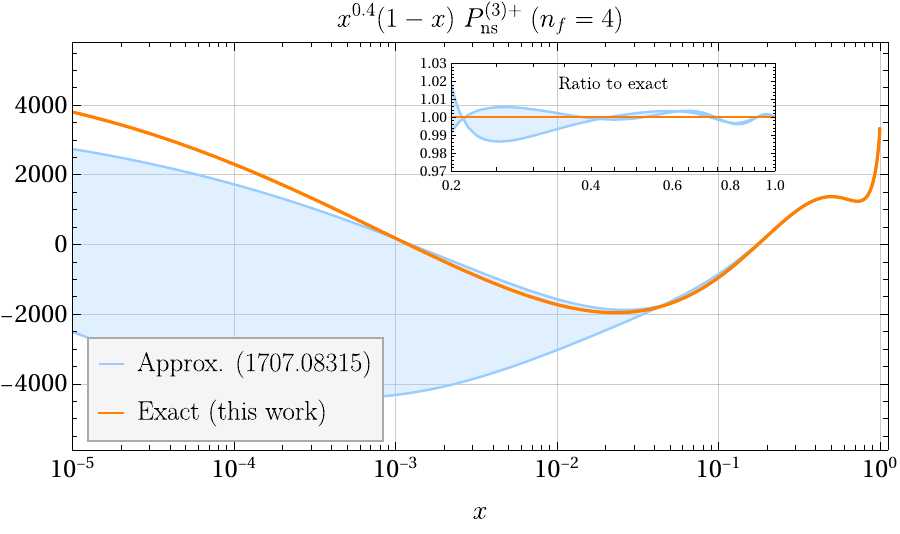} ~
  \includegraphics[width=0.49\linewidth]{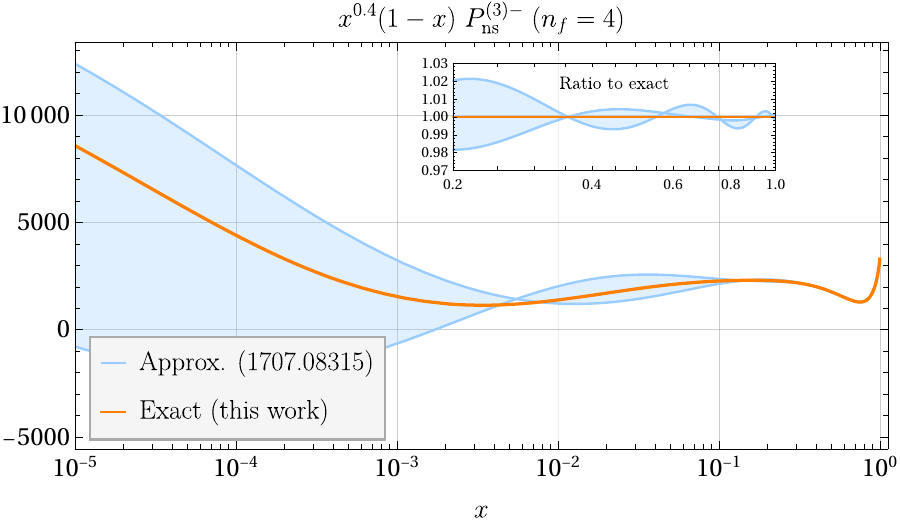}\\[1em]
  \includegraphics[width=0.49\linewidth]{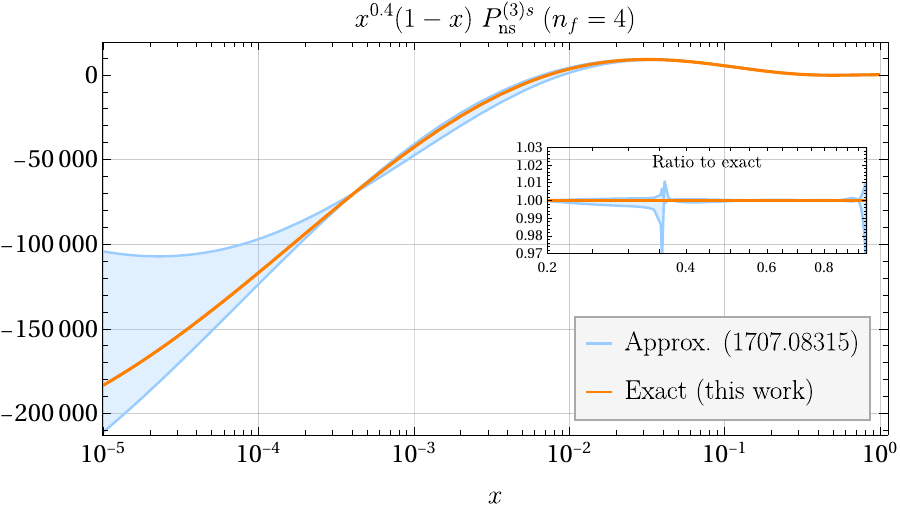}
  \caption{
    The four-loop non-singlet splitting functions $P^{(3)+}_\mathrm{ns}$, $P^{(3)-}_\mathrm{ns}$ and $P^{(3)s}_\mathrm{ns}$, evaluated for $\nf=4$ and compared to approximations from ref.~\cite{Moch:2017uml}.
    All functions are multiplied by the factor $x^{0.4}(1-x)$ for presentation.
  }
  \label{fig:P3results}
\end{figure*} 

The coefficients $C_4$ and $D_4$ in eq.~\eqref{eq:xto1limit} are 
related to the lower-order cusp anomalous dimension, virtual 
anomalous dimension, and $\beta$-function coefficients through an 
all-order conjecture established in~\cite{Dokshitzer:2005bf,Basso:2006nk}. The explicit relations at four loops read
\begin{align}
    C_4 = 2 A_1 A_3 + A_2^2 \,, \quad
    D_4 = \sum_{l=1}^{3} A_l \left( B_{4-l} - \beta_{3-l} \right) \,.
\end{align}
Our results verify this conjecture at this loop order. 

Our analytical results also allow us to extract the
behavior of the non-singlet splitting functions
in the limit $x \to 0$, stated in the appendix for $N_C=3$.
We observe that 
the logarithmically enhanced terms proportional to $\log^k x$ with $k=6,5$ for $P^{(3)+}_{\text{ns}}$ agree with the expressions given in~\cite{Davies:2022ofz}.
For $k=4$, we find deviations proportional to $\zeta_2$ 
in all five non-$\nf$ color structures, which conspire to 
cancel in
the leading-color limit.

In \cref{fig:P3results} we compare the non-distribution part
(defined by omitting the terms proportional to $\delta(1-x)$ and arising from the 
subtraction at $x=1$ originating from the plus distribution)
of the non-singlet splitting functions with $\nf=4$
with the approximations provided in the \texttt{Fortran}
code of ref.~\cite{Moch:2017uml}.
We observe that overall the approximations agree well with the exact results at the level of the quoted uncertainties.
Only $P_{\mathrm{ns}}^{(3)+}$ in the low-$x$ region is slightly above the approximation's envelope.

Our analytic results allow us to construct simple power-logarithmic approximations which contain only polynomials in $\{x, \log x, \log (1-x), \delta(1-x), \left[\frac{1}{1-x}\right]_+\}$. The constructed results have an accuracy better than $10^{-6}$ over the whole $x$ range of $x\in (0,1)$ and can be used for more convenient numerical applications.

Finally, we used the correspondence~\cite{Moch:2017uml} between space-like and time-like non-singlet anomalous dimensions, derived from reciprocity relations \cite{Basso:2006nk,Dokshitzer:2006nm,Chen:2020uvt}, to 
determine the four-loop non-singlet time-like splitting functions, which are relevant to the 
scale evolution of  fragmentation functions. 

We provide all our analytical results as well
as our numerical approximations in the ancillary
files.

\section{Summary and Outlook}

In this letter, we derived the four-loop non-singlet 
splitting functions in QCD in analytical form, based 
on a calculation of the anomalous dimensions of the non-singlet operators 
for symbolic Mellin moment $n$. We provide results for 
both space-like and time-like kinematics. 

Our calculation 
confirms earlier partial results for the four-loop contributions 
involving closed fermion loops 
proportional to $\nf^2$~\cite{Davies:2016jie}, $\nf$~\cite{Gehrmann:2023iah,Kniehl:2025ttz} and in the planar 
limit of QCD~\cite{Moch:2017uml}, as well as results 
for fixed values of $n\leq 16$
for the non-$\nf$ terms~\cite{Moch:2017uml,Moch:2021qrk} at four loops. 
Previously available approximations~\cite{Moch:2017uml} to the 
four-loop non-singlet splitting functions were 
based on the 
fixed-moment results. With the complete results obtained here, we show that these approximations are reliable within their 
quoted uncertainty band.
Our newly derived results significantly reduce the
uncertainties of the four-loop non-singlet splitting functions,
in particular for smaller values of $x$.

The analytic expressions can be used to study
the limiting behavior of the four-loop non-singlet splitting functions at $x\to 1$ and $x\to 0$, enabling
systematic control not only on the logarithmically enhanced terms, but also 
on power-suppressed contributions. 
From the $x\to 1$ behavior, we obtained  exact values for the 
four-loop virtual and rapidity anomalous dimensions, which enter 
resummation at N\textsuperscript{4}LL.

\emph{Acknowledgments.}
We thank Andreas Vogt for enlightening discussions on the structure and implications of the results computed here.
This work has been supported by the European Research Council (ERC) 
under the European Union's Horizon 2020 research and innovation programme grant agreement 101019620 (ERC Advanced Grant TOPUP).

\bibliography{letter}

\appendix

\begin{widetext}

\section{Four-loop rapidity anomalous dimension}
\label{app:adm}
Our result for the virtual anomalous dimension allows one to determine the analytical values of constants in the four-loop rapidity anomalous dimension that had previously been known only through numerical approximations.
Following the conventions of \cite{Moult:2022xzt},  the fully analytical result for the rapidity anomalous dimension for representation $r = \{F,A\}$ reads
\begin{align}
  \nonumber
  \gamma_3^r = \;
  & \CS{C_A^3 C_r} \Bigg\{
  \frac{9671}{12} \zeta_7
  +216 \zeta_5 \zeta_2
  +355 \zeta_4 \zeta_3
  -\frac{20933}{36} \zeta_6
  -\frac{5027}{9} \zeta_3^2
  -\frac{45031}{27} \zeta_5 
  -\frac{6190}{9} \zeta_3 \zeta_2
  +\frac{2068}{9} \zeta_4
  \\ \nonumber & 
  +\frac{301822}{81} \zeta_3
  +\frac{395455}{486} \zeta_2
  -\frac{28325071}{8748}
  \Bigg\}
  \\\nonumber
  +\; &\CS{\frac{d_A^{abcd}d_r^{abcd}}{n_r}} \left(
  1058 \zeta_7
  -192 \zeta_5 \zeta_2
  +120 \zeta_4 \zeta_3
  -\frac{1100}{3} \zeta_6
  -\frac{440}{3} \zeta_3^2
  -\frac{2680}{9} \zeta_5
  -8 \zeta_4
  +\frac{208}{9} \zeta_3
  + 48 \zeta_2
  \right)
  \\\nonumber
  +\; &\CS{\nf C_A^2 C_r} \left(
  \frac{247}{18} \zeta_6
  -\frac{850}{9} \zeta_3^2
  +\frac{1138}{27} \zeta_5
  +\frac{404}{9} \zeta_3 \zeta_2
  -\frac{526}{9} \zeta_4
  -\frac{32018}{81} \zeta_3
  -\frac{41047}{243} \zeta_2
  +\frac{11551831}{11664}
  \right)
  \\\nonumber
  +\; &\CS{\nf C_A C_F C_r}  \left(
   -8 \zeta_6
   +156 \zeta_3^2
   -\frac{652}{3} \zeta_5
  +\frac{152}{3} \zeta_3 \zeta_2
  -100 \zeta_4
  -\frac{16819}{27} \zeta_3
  -\frac{1741}{18} \zeta_2
  +\frac{1870013}{1944}
   \right)
  \\\nonumber
  +\; &\CS{\nf C_F^2 C_r}  \left(
   100 \zeta_6
   +40 \zeta_3^2
   +\frac{800}{3} \zeta_5
  -37 \zeta_4
  -\frac{2212}{9} \zeta_3
  -\frac{21037}{216}
   \right)
  \\\nonumber
  +\; &\CS{\nf \frac{d_F^{abcd}d_r^{abcd}}{n_r}} \left(
  \frac{400}{3} \zeta_6
  +\frac{160}{3} \zeta_3^2
  +\frac{800}{9} \zeta_5
  +16 \zeta_4
  -\frac{320}{9} \zeta_3
  -128 \zeta_2
  \right)
  \\\nonumber
  +\; &\CS{\nf^2 C_A C_r} 
   \left(
   \frac{368}{9}\zeta_5
   +\frac{56}{9} \zeta_3 \zeta_2
   +\frac{40}{9} \zeta_4
   -\frac{5564}{81} \zeta_3
   +\frac{1688}{243} \zeta_2
   -\frac{898033}{11664}
   \right)
  \\\nonumber
  +\; &\CS{\nf^2 C_F C_r} \left(
  8 \zeta_5
  +\frac{40}{3} \zeta_4
  +\frac{1732}{27} \zeta_3
  -\frac{110059}{972}
  \right)\\
  +\; &\CS{\nf^3 C_r} \left(
  -\frac{4}{9} \zeta_4
  +\frac{40}{9} \zeta_3
  +\frac{2608}{2187}
  \right)\nonumber
\end{align}
where $n_r$ is the dimension of the representation with $n_r = N_c$ for the fundamental representation and $n_r = N_c^2-1$ for the adjoint representation.

\section{Splitting functions in the small-$x$ limit}
\label{app:smallx}

To derive the limit $x \to 0$ of the non-singlet splitting functions, we expand harmonic polylogarithms and rational functions in $x$.
In QCD with $N_c=3$ we obtain
\begin{align}
P_{\mathrm{ns}}^{(3)+} &= \frac{256}{729} \log^6 x
  + \left(\frac{3200}{243} - \frac{256}{243}\,\nf\right) \log^5 x
  + \left(
   - \frac{14240}{243}\zeta_2 
  + \frac{89008}{243}
    - \frac{7616}{243}\,\nf + \frac{64}{81}\,\nf^2\right) \log^4 x
\nonumber\\&\quad
  + \Bigg\{
   \frac{1024}{243}\zeta_3
  - \frac{85504}{81}\zeta_2
  +\frac{1062692}{243} 
     + \left( \frac{2368}{27}\zeta_2-\frac{43208}{81}\right)\nf
    + \frac{4016}{243}\,\nf^2 - \frac{32}{243}\,\nf^3\Bigg\} \log^3 x
\nonumber\\&\quad
  + \Bigg\{
      \frac{362560}{81}\zeta_4
      + \frac{55808}{81}\zeta_3
      - \frac{1181128}{81}\zeta_2
      + \frac{2763724}{81} 
    + \left( - \frac{11264}{81}\zeta_3
    + \frac{82096}{81}\zeta_2
    -\frac{1010488}{243} 
   \right)\nf
\nonumber\\&\quad
  {}+ \left( - \frac{1472}{81}\zeta_2+\frac{103856}{729}\right)\nf^2
    - \frac{352}{243}\,\nf^3\Bigg\} \log^2 x
  + \Bigg\{
  - \frac{124160}{27}\zeta_5
  + \frac{59456}{27}\zeta_3\zeta_2
    + \frac{1222496}{81}\zeta_4
\nonumber\\&\quad
    + \frac{479584}{81}\zeta_3
  - \frac{3824144}{81}\zeta_2
  +\frac{8777560}{81} 
    + \left(-\frac{31904}{27}\zeta_4
    - \frac{68480}{81}\zeta_3
    + \frac{429088}{81}\zeta_2
    -\frac{3673924}{243}
    \right)\nf
\nonumber\\&\quad
  {}+ \left(
   \frac{128}{3}\zeta_3
    - \frac{9920}{81}\zeta_2
    +\frac{395972}{729}
    \right)\nf^2
    - \frac{256}{81}\,\nf^3\Bigg\} \log x 
    +  \frac{1912120}{243} \zeta_6
    +\frac{290320}{81}\zeta_3^2
    -\frac{2676784}{243}\zeta_5
    \nonumber\\&\quad
    +\frac{142720}{81}\zeta_3 \zeta_2
    +\frac{3251956}{81} \zeta_4
    -\frac{19373648}{243}\zeta_2
   +\frac{1234960}{243}\zeta_3
   +\frac{39538148}{243}
   +\Bigg\{
   -\frac{11408}{9}\zeta_6
   \nonumber\\&\quad
   -\frac{2944}{9}\zeta_3^2
   +\frac{160352}{81}\zeta_5
   +\frac{35392}{81}\zeta_3 \zeta_2
   -\frac{116504}{81} \zeta_4
   -\frac{356320}{81}\zeta_3
   +\frac{1883392}{243}\zeta_2
   -\frac{5427338}{243}
   \Bigg\} \nf \nonumber\\&\quad +\Bigg\{
   -\frac{2528}{81}\zeta_4
   +\frac{40448}{243}\zeta_3
   -\frac{83168}{729}\zeta_2
   +\frac{430894}{729}
   \Bigg\} n_f^2+\left(\frac{128}{81}\zeta_3-\frac{256}{243}\right) n_f^3 + \mathcal{O}(x),  
\\
\nonumber \\ 
P_{\mathrm{ns}
}^{(3)-} &= \frac{3632}{3645} \log^6 x
  + \left(\frac{22496}{1215} - \frac{1184}{1215}\,\nf\right) \log^5 x
  + \left(- \frac{21104}{243}\zeta_2+\frac{107168}{243} 
    - \frac{8408}{243}\,\nf + \frac{80}{81}\,\nf^2\right) \log^4 x \nonumber\\&\quad
  + \Bigg\{
  - \frac{16384}{243}\zeta_3
  - \frac{104960}{81}\zeta_2 
  +\frac{1318516}{243}
  + \left(\frac{7744}{81}\zeta_2 -\frac{50888}{81}\right)\nf
    + \frac{14288}{729}\,\nf^2 - \frac{32}{243}\,\nf^3\Bigg\} \log^3 x\nonumber\\&\quad
  + \Bigg\{  \frac{289040}{81}\zeta_4 
  + \frac{2752}{9}\zeta_3
  +\frac{3326836}{81}
  - \frac{140024}{9}\zeta_2
    + \left(- \frac{17600}{81}\zeta_3
    + \frac{81488}{81}\zeta_2
    - \frac{14072}{3} 
    \right)\nf \nonumber\\&\quad
    + \left(
    - \frac{448}{27}\zeta_2
    +\frac{37808}{243}
    \right)\nf^2
    - \frac{352}{243}\,\nf^3\Bigg\}\log^2 x
  + \Bigg\{
    - \frac{899296}{81}\zeta_5 
    + \frac{104960}{81}\zeta_3\zeta_2
    + \frac{2944304}{243}\zeta_4
    \nonumber\\&\quad 
    + \frac{705760}{81}\zeta_3
    - \frac{11387728}{243}\zeta_2
    + \frac{28821224}{243}
  + \left(
    - \frac{3712}{3}\zeta_4
    - \frac{35008}{27}\zeta_3
    + \frac{131936}{27}\zeta_2
    -\frac{3736868}{243}
    \right)\nf \nonumber\\&\quad
    + \left(
     \frac{3584}{81}\zeta_3
    - \frac{27328}{243}\zeta_2
    +\frac{415300}{729}
    \right)\nf^2
    - \frac{256}{81}\,\nf^3\Bigg\}\log x
    +\frac{45400}{81}\zeta_6
    +\frac{325360}{81} \zeta_3^2
    +\frac{45184}{81} \zeta_3 \zeta_2 
    \nonumber\\&\quad
    -\frac{251216}{9} \zeta_5
    +\frac{4490708}{81} \zeta_4
    +\frac{174592}{9} \zeta_3
    -\frac{22420288}{243} \zeta_2
   +\frac{14570924}{81}
   + \Bigg\{
   -\frac{11408}{9} \zeta_6
   -\frac{2944}{9} \zeta_3^2
   \nonumber\\&\quad
   +\frac{20512}{9} \zeta_5
   +\frac{27968}{81} \zeta_3 \zeta_2
   -\frac{181400}{81} \zeta_4
   -\frac{488320}{81} \zeta_3
   +\frac{1955552}{243} \zeta_2
   -\frac{5270602}{243}
   \Bigg\} n_f  \nonumber\\&\quad
   +  \left(
   -32 \zeta_4
   +\frac{45824}{243} \zeta_3
   -\frac{80992}{729} \zeta_2
   +\frac{448622}{729}\right) n_f^2
   + \left(\frac{128}{81} \zeta_3
   -\frac{256}{243}\right) n_f^3+  \mathcal{O}(x),
 \\ 
 \nonumber \\ 
\frac{P^{(3)s}_{\mathrm{ns}}}{\nf} &= -\frac{152}{81} \log^6 x
  + \left(\frac{128}{81} - \frac{32}{81}\,\nf\right) \log^5 x
  + \left(\frac{3640}{81}\zeta_2 -\frac{7360}{81} 
    - \frac{160}{27}\,\nf\right) \log^4 x
  + \Bigg\{\frac{40000}{81}\zeta_3 \nonumber\\&\quad
    + \frac{9920}{81}\zeta_2 
    - \frac{77600}{81}
    + \left(\frac{320}{81}\zeta_2-\frac{320}{27} \right)\nf
    \Bigg\}\log^3 x
  + \Bigg\{ 
   \frac{88760}{27}\zeta_4 
  + \frac{88480}{27}\zeta_3
  - \frac{2720}{27}\zeta_2
  \nonumber\\&\quad
  - \frac{135680}{27}
    + \left(- \frac{640}{9}\zeta_3
    + \frac{640}{3}\zeta_2
    -\frac{8960}{27}
    \right)\nf
    \Bigg\}\log^2 x
  + \Bigg\{
    + \frac{86080}{9}\zeta_5
    - \frac{33920}{9}\zeta_3\zeta_2
  + \frac{19600}{9}\zeta_4 
  \nonumber\\&\quad
  + \frac{267200}{27}\zeta_3
  + \frac{298240}{27}\zeta_2
  -\frac{733280}{27}
    + \left(
    - \frac{800}{27}\zeta_4
    + \frac{10240}{27}\zeta_3
    - \frac{4160}{27}\zeta_2
    -\frac{7360}{9}
    \right)\nf
    \Bigg\}\log x \nonumber\\&\quad
    - \frac{93500}{81}\zeta_6
   - \frac{33520}{9}\zeta_3^2 
    + \frac{213760}{27}\zeta_5
    - \frac{283840}{27}\zeta_3\zeta_2
   - \frac{672320}{9}
    + \frac{461120}{27}\zeta_4
    + \frac{1031920}{27}\zeta_3
  + \frac{622720}{27}\zeta_2
\nonumber\\&\quad
    + \left(
    - \frac{12160}{27}\zeta_5
    + \frac{3200}{9}\zeta_3\zeta_2
    - \frac{4160}{27}\zeta_4
    - \frac{8000}{9}\zeta_3
    + \frac{17440}{27}\zeta_2
    -\frac{33920}{27}
    \right)\nf + \mathcal{O}(x)\,.
\end{align}
The results for a generic gauge group can be found in the ancillary files.

\end{widetext}

\end{document}